\documentclass[prb,twocolumn,showpacs]{revtex4}
\usepackage{amsmath}
\usepackage{dcolumn}
\usepackage{graphicx}

\begin{document}

\title[Short title for running header]{Orbital magnetic response and the anisotropy of magnetic susceptibility in the Iron-based superconductors}
\author{Yuehua Su$^{1}$ and Tao Li$^{2}$}
\affiliation{$^{1}$Department of Physics, Yantai University,
Yantai 264005, P.R.China\\ $^{2}$Department of Physics, Renmin
University of China, Beijing 100872, P.R.China}
\date{\today}

\begin{abstract}
We propose that the orbital angular momentum of the conduction
electrons in the Iron-based superconductors is activated in their
low energy physics. Using a five-band tight-binding model derived
from fitting the LDA band structure, we find that the orbital
magnetic susceptibility of the conduction electrons in such a
multi-orbital system is several times larger than the Pauli spin
susceptibility and is comparable in magnitude to the observed total
magnetic susceptibility. The orbital magnetic susceptibility in the
Fe-As plane($\chi^{x}_{L}$) is found to be larger than that
perpendicular to the Fe-As plane($\chi^{z}_{L}$) by a factor about
two and the total magnetic susceptibility in the normal state can be
fitted with formula $\chi(T,\theta)\approx
\chi_{s}(T)+\chi_{L}(\theta)$, where $\chi_{s}(T)$ is the
temperature dependent isotropic part due to spin and
$\chi_{L}(\theta)$ is the temperature independent anisotropic part
due to orbital. In the superconducting state, $\chi^{x}_{L}$ is
found to be significantly reduced as the pairing gap develops, while
$\chi^{z}_{L}$ is almost not affected by the superconducting
transition. We argue the large anisotropy observed in the bulk
magnetic susceptibility and the Knight shift in the Iron-based
superconductors should be attributed to the orbital magnetic
response of their conduction electrons.
\end{abstract}


\maketitle

One special feature of the newly discovered Iron-based
superconductors\cite{Hosono,Rotter,CQJin,MKWu} is their
multi-orbital nature. In conventional superconductors, in which only
one orbital plays an essential role around the Fermi energy, the
orbital angular momentum is quenched either as a result of the
s-wave character of the orbital, or by the crystal field splitting
effect. However, from LDA band structure calculation, it is reported
that all the five Fe 3d orbital play essential role in forming the
low energy degree of freedom around the Fermi energy and the crystal
field splitting is extremely small. This situation has caused great
complexities in the model study. However, it also generates the
interesting opportunity to explore the physics of the orbital degree
of freedom in this system. The orbital character near the Fermi
energy has now been extensively studied by angle-resolved
photonemission spectroscopy (ARPES) and the LDA result is to a large
extent confirmed.\cite{Hasan, Shimojima, Feng, ZXShen}

Many novel properties of the Iron-based superconductors have been
attributed to their multi-orbital nature. For example, the intimate
relation between the structural and magnetic phase
transitions\cite{Dai,Huang} has been proposed to originate from an
orbital-related mechanism,\cite{CCLee,Phillips}. The same picture
also provides a reasonable interpretation for the unusual in-plane
anisotropy in resistivity\cite{ChuPRB,ChuScience} and the $d_{xz}$
and $d_{yz}$ band splitting observed in the ARPES
measurement.\cite{ZXShen} In these situations, the multi-orbital
nature manifests itself in the form of a static structure, a more
interesting possibility is that the orbital degree of freedom
appearing as a dynamic mode in the low energy physics, contributing
to various kinds of response and relaxation processes. The purpose
of this paper is to investigate the contribution of the orbital
angular momentum of the conduction electrons to the magnetic
susceptibility.

The magnetic susceptibility of a metal is usually attributed to the
response of its spin degree of freedom, which in the absence of the
spin-orbital coupling is isotropic. The orbital magnetic
susceptibility, which is in principle anisotropic, is usually small
as a result of the quenching of the orbital angular momentum of the
conduction electrons. The small remnant orbital magnetic response,
namely the well known Van Velck paramagnetism, is controlled by the
large gap separating the occupied bands from the unoccupied atomic
levels that carry orbital angular momentum and is expected to show
negligible temperature dependence. However, if the conduction
electron itself carries orbital angular momentum, the orbital
magnetic response of the system would be much larger and be
sensitive to the changes of electronic state of the system.

\begin{figure}[htp]
\includegraphics[width=8.5cm,angle=0]{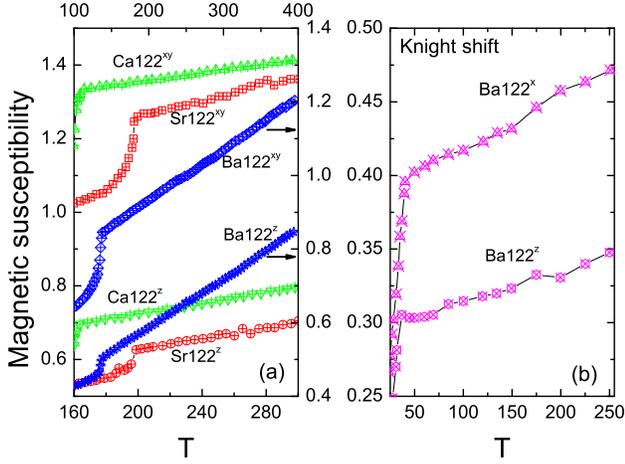}
\caption{The experimental results of the magnetic susceptibility(a)
and Knight shift(b) for some Iron-based supercondutors. The magnetic
susceptibility data of Ba122 system are quoted from Ref.
\onlinecite{ChenBa}, data Ca122 system from Ref.
\onlinecite{ChenCa}(both in unit of 10$^{-3}$ emu/Oe), and data of
Sr122 system from Ref. \onlinecite{CanfieldSr}(in unit of $10^{-3}$
emu/mol). The Knight shift data are quoted from Ref.
\onlinecite{GQZheng}. Here the superscripts $xy/x$ and $z$ denote
the direction of the applied magnetic field. } \label{fig1}
\end{figure}

Measurement of the magnetic susceptibility on the Iron-based
superconductor has produced several intriguing results. Firstly, the
magnetic susceptibility is found to show significant temperature
dependence.\cite{ChenBa,ChenCa,CanfieldSr,Klingeler1111,GQZheng}
Such a temperature dependence is unexpected for a weakly correlated
system with a large band width. In the literature, this unusual
temperature dependence is either attributed to the strong
correlation effect, or to the proximity of the system to a semimetal
phase\cite{GMZhang,SPKou,Chubukov}. Secondly, the magnetic
susceptibility is found to be strongly
anisotropic.\cite{ChenCa,ChenBa,CanfieldSr,Klingeler1111,GQZheng} As
shown in Fig. \ref{fig1}, the susceptibility in the Fe-As plane is
much larger than that perpendicular to the Fe-As plane and their
difference is nearly temperature independent. An understanding of
such anisotropy in the magnetic susceptibility is still absent and
we will show that it can be originated from the contribution of the
orbital angular momentum of the conduction electrons.

In this paper, we study the orbital magnetic response of the
conduction electrons in the Iron-based superconductors. It is found
that the orbital magnetic susceptibility of the conduction electrons
is several times larger than the Pauli spin susceptibility and is
comparable in magnitude with the observed total magnetic
susceptibility. The orbital magnetic susceptibility in the Fe-As
plane($\chi^{x}_{L}$) is found to be larger than the susceptibility
perpendicular to the Fe-As plane($\chi^{z}_{L}$) by a factor about
2. In the normal state, the magnetic susceptibility can be separated
into a temperature dependent isotropic part $\chi_{s}(T)$ due to
spin response and a temperature independent anisotropic part
$\chi_{L}(\theta)$ due to orbital response, where $\theta$ is angle
between the magnetic field and the normal of the Fe-As plane. In the
superconducting state, when a $s^{\pm}$-wave intraband pairing is
assumed, $\chi^{x}_{L}$ is found to be significantly reduced as the
pairing gap develops, while $\chi^{z}_{L}$ is almost unaffected.
Unlike the Pauli spin susceptibility, the orbital magnetic
susceptibility is nonzero in the zero temperature limit as a result
of the interband contribution to the orbital magnetic response.

The properties of the Iron-based superconductors are sensitive to
model parameters. To understand its physics, a realistic band
structure is indispensable. In this paper, we will adopt the
five-band tight-binding model derived from fitting the LDA band
structure\cite{Kuroki}. The model reads (following the notations of
Ref. \onlinecite{Kuroki}),
\begin{eqnarray}
 H_{0}&=&\sum_{i,j}\sum_{\mu,\nu,\sigma}[t(x_{i}-x_{j},y_{i}-y_{j};\mu,\nu)c_{i,\mu,\sigma}^{\dagger}c_{j,\nu,\sigma} \label{eqn1} \\
 &+&t(x_{j}-x_{i},y_{j}-y_{i};\nu,\mu)c_{j,\nu,\sigma}^{\dagger}c_{i,\mu,\sigma}]
 +\sum_{i,\mu,\sigma}\varepsilon_{\mu}n_{i,\mu,\sigma} , \nonumber
\end{eqnarray}
where $\mu,\nu=1,2,3,4,5$ denote the five Fe 3d orbital
$3d_{3z^{2}-r^{2}}$, $3d_{xz}$, $3d_{yz}$, $3d_{xy}$ and
$3d_{x^{2}-y^{2}}$ respectively. $t(\Delta x,\Delta y;\mu,\nu)$
denotes the in-plane hopping integral between the $\mu$-th and
$\nu$-th orbitals at the lattice distance $(\Delta x, \Delta y)$,
$\varepsilon_{\mu}$ is the on-site energy of the $\mu$-th orbital.
The values of these model parameters are given in
Ref.\onlinecite{Kuroki}. The model Hamiltonian Eq.(\ref{eqn1}) can
be diagonalized in momentum space, in which it takes the form
\begin{equation}
H_{0}=\sum_{{\mathbf k},\alpha,\sigma}\epsilon_{{\mathbf
k},\alpha}c^{\dagger}_{k,\alpha,\sigma}c_{{\mathbf
k},\alpha,\sigma} , \label{eqn2}
\end{equation}
where $\epsilon_{{\mathbf k},\alpha}$ is the band energy for the
$\alpha$-th band.

To describe the superconducting state, we model the pairing
potential with a sign-changing $s$-wave form
$\Delta_{\alpha\beta}({\mathbf
k},T)=\Delta(T)\delta_{\alpha\beta}\cos(k_{x})\cos(k_{y})$. Note
that only the intraband pairing is considered. Such a pairing
symmetry is consistent with the weak coupling
spin-fluctuation-exchange mechanism\cite{Kuroki} or the strong
coupling superexchange mechanism\cite{JPHuPair} for the
superconductivity in the Iron-based superconductors. However, as
will be clear below, the detailed form of the pairing potential
(except its intraband pairing nature) is not essential for the
uniform susceptibility. So we will take the pairing potential
adopted only as a simplified way to induce a full gap on the Fermi
surfaces of the system.

The orbital magnetic susceptibility is defined through the
correlation function of the orbital magnetic moment in the following
way
\begin{equation}
\chi^{a}_{L}({\mathbf q},\tau)=-(g_{L}\mu_{B})^2\langle
T_{\tau}L^{a}({\mathbf q},\tau)L^{a}(-{\mathbf q},0) \rangle ,
\label{eqn3}
\end{equation}
in which $L^{a}({\mathbf q},\tau)$ denotes the Fourier component of
the orbital magnetic moment density in the $a$-direction. The
orbital magnetic moment on a given site $i$ is defined as
$L^{a}_{i}=\sum_{\nu,\nu',s}c^{\dagger}_{i,\nu,s}l^{a}_{\nu,\nu'}c_{i,\nu',s}$,
where $l^{a}_{\nu,\nu'}$ is the matrix element of the orbital
magnetic moment in the space spanned by the five orbital
$3d_{3z^{2}-r^{2}}$, $3d_{xz}$, $3d_{yz}$, $3d_{xy}$ and
$3d_{x^{2}-y^{2}}$. Here we set $g_{L}\mu_{B}=1$.

To derive an expression for the matrix element $l^{a}_{\nu,\nu'}$,
we approximate the Wannier functions with the Fe 3d atomic orbital.
We then have
\begin{eqnarray}
\left(\begin{array}{c}
        |1\rangle \\
        |2 \rangle \\
        |3 \rangle \\
        |4 \rangle \\
        |5\rangle \\
      \end{array}\right)=\frac{1}{\sqrt{2}}\left(\begin{array}{ccccc}
                    0 & 0 & \sqrt{2} & 0 & 0 \\
                    0 & 1 & 0 & 1 & 0 \\
                    0 & -i & 0 & i & 0 \\
                    -i & 0 & 0 & 0 & i \\
                    1 & 0 & 0 & 0 & 1
                  \end{array}
\right)\left(\begin{array}{c}
        |2,2 \rangle \\
        |2,1 \rangle \\
        |2,0 \rangle \\
        |2,-1\rangle \\
        |2,-2\rangle \\
      \end{array}\right) , \label{eqn4}
\end{eqnarray}
where $|\mu\rangle$ ($\mu=1,2,3,4,5$) denote the five Wannier
functions, $|l,m\rangle\propto Y^{m}_{l}$ denotes the eigenstate of
the orbital angular momentum $(l^{2},l^{z})$, whose eigenvalue for
$l^{z}$ is $m$. With the transformation Eq.(\ref{eqn4}), the matrix
$l^{z}_{\nu,\nu'}$ and $l^{x}_{\nu,\nu'}$ can be found as
\begin{eqnarray}
l^{z}=\left(\begin{array}{ccccc}
                    0 & 0 & 0 & 0 & 0 \\
                    0 & 0 & i & 0 & 0 \\
                    0 & -i & 0 & 0 & 0 \\
                    0 & 0 & 0 & 0 & 2i \\
                    0 & 0 & 0 & -2i & 0
                  \end{array}
\right) \nonumber
\end{eqnarray}
and
\begin{eqnarray}
l^{x}=\left(\begin{array}{ccccc}
                    0 & 0 & -i\sqrt{3} & 0 & 0 \\
                    0 & 0 & 0 & 0 & -i \\
                    i\sqrt{3} & 0 & 0 & i & 0 \\
                    0 & 0 & -i & 0 & 0 \\
                    0 & i & 0 & 0 & 0
                  \end{array}
\right).\nonumber
\end{eqnarray}

The orbital magnetic susceptibility is readily obtained as follows
\begin{eqnarray}
\chi^{a}_{L}(T)&=&\lim_{{\mathbf q}\rightarrow
0}\frac{1}{N}\sum_{{\mathbf
k},\alpha,\beta}\left[\frac{f(E_{\mathbf{
k+q},\beta})-f(E_{{\mathbf k},\alpha})}
{E_{{\mathbf k},\alpha}-E_{\mathbf{ k+q},\beta}} \right.\label{eqn5}\\
&\times&\left(1+\frac{\xi_{{\mathbf k},\alpha}\xi_{\mathbf{
k+q},\beta}+\Delta_{{\mathbf k},\alpha}\Delta_{\mathbf{
k+q},\beta}}{E_{{\mathbf k},\alpha}E_{\mathbf{
k+q},\beta}}\right)\times
\left|O^{a}_{{\mathbf k},\alpha,\beta}\right|^{2} \nonumber \\
&+&\frac{1-f(E_{\mathbf{ k+q},\beta})-f(E_{{\mathbf
k},\alpha})}{E_{{\mathbf k},\alpha}+E_{\mathbf{ k+q},\beta}} \nonumber \\
&\times&\left. \left(1-\frac{\xi_{{\mathbf k},\alpha}\xi_{\mathbf{
k+q},\beta}+\Delta_{{\mathbf k},\alpha}\Delta_{\mathbf{
k+q},\beta}}{E_{{\mathbf k},\alpha}E_{\mathbf{
k+q},\beta}}\right)\times \left|O^{a}_{{\mathbf
k},\alpha,\beta}\right|^{2} \right]. \nonumber
\end{eqnarray}
Here $E_{{\mathbf k},\alpha}$ denotes the excitation energy of the
quasiparticle in the $\alpha$-th band and is given by $E_{{\mathbf
k},\alpha}=\sqrt{\xi^{2}_{{\mathbf k},\alpha}+\Delta^{2}_{{\mathbf
k},\alpha}}$. $\xi_{{\mathbf k},\alpha}=\epsilon_{{\mathbf
k},\alpha}-\mu$ and $\Delta_{{\mathbf k},\alpha}$ are the band
energy and the pairing gap of the $\alpha$-th band, $\mu$ is the
chemical potential. $O^{a}_{{\mathbf
k},\alpha,\beta}=\sum_{\nu,\nu'}U^{*}_{{\mathbf
k},\nu,\alpha}l^{a}_{\nu,\nu'}U_{{\mathbf k},\nu',\beta}$ is the
matrix element of $l^{a}$ in the basis of the band eigenstate at
momentum $\mathbf{k}$. $U_{{\mathbf k},\nu,\alpha}$ is the
$\alpha$-th eigenvector of the band Hamiltonian at momentum
${\mathbf k}$. In deriving Eq.(\ref{eqn5}), we have used the
inversion symmetry of the system and the fact that
$l^{a}_{\nu,\nu'}=-l^{a}_{\nu',\nu}$.

For comparison, we also calculate the Pauli spin susceptibility of
the band electrons. The Pauli susceptibility is defined through the
following spin-spin correlation function,
\begin{equation}
\chi_{S}({\mathbf q},\tau)=-(g_{S}\mu_{B})^2\langle
T_{\tau}S^{z}({\mathbf q},\tau)S^{z}(-{\mathbf q},0) \rangle ,
\label{eqn6}
\end{equation}
where $g_{S}\mu_{B}=2$ and the spin density operator at lattice
site $i$ can be written as
$S^{z}_{i}=\sum_{\mu,s,s'}c^{\dagger}_{i,\mu,s}\sigma^{z}_{s,s'}c_{i,\mu,s'}$.
The uniform bare spin susceptibility can be shown to be given by
\begin{eqnarray}
\chi_{S}(T)&=&\lim_{{\mathbf q}\rightarrow
0}\frac{1}{N}\sum_{{\mathbf k},\alpha}\left[\frac{f(E_{\mathbf{
k+q},\alpha})-f(E_{{\mathbf k},\alpha})}{E_{{\mathbf
k},\alpha}-E_{\mathbf{ k+q},\alpha}}\right.
\label{eqn7}\\
&\times&\left(1+\frac{\xi_{{\mathbf k},\alpha}\xi_{\mathbf{
k+q},\alpha}+\Delta_{{\mathbf k},\alpha}\Delta_{\mathbf{
k+q},\alpha}}{E_{{\mathbf k},\alpha}E_{\mathbf{
k+q},\alpha}}\right)
\nonumber \\
&+&\frac{1-f(E_{\mathbf{ k+q},\alpha})-f(E_{{\mathbf
k},\alpha})}{E_{{\mathbf k},\alpha}+E_{\mathbf{ k+q},\alpha}}
\nonumber \\
&\times&\left.\left(1-\frac{\xi_{{\mathbf k},\alpha}\xi_{\mathbf{
k+q},\alpha}+\Delta_{{\mathbf k},\alpha}\Delta_{\mathbf{
k+q},\alpha}}{E_{{\mathbf k},\alpha}E_{\mathbf{
k+q},\alpha}}\right) \right] . \nonumber
\end{eqnarray}

The difference between Eq.(\ref{eqn5}) and Eq.(\ref{eqn7}) lies in
the fact that the orbital magnetic response has contribution from
both the intraband and interband processes, while the spin
susceptibility has only intraband contribution, the reason for the
latter is that the spin density operator is diagonal in the
orbital space. Such a difference will have important consequence
in the superconducting state. As the coherence factor,
$(1-\frac{\xi_{{\mathbf k},\alpha}\xi_{\mathbf{
k+q},\alpha}+\Delta_{{\mathbf k},\alpha}\Delta_{\mathbf{
k+q},\alpha}}{E_{{\mathbf k},\alpha}E_{\mathbf{ k+q},\alpha}})$,
vanishes when $q\rightarrow 0$ in the superconducting state,
$\chi_{S}(T)$ will show activation behavior if there is a full gap
on the Fermi surface. On the other hand, the coherence factor for
the orbital magnetic susceptibility, $(1-\frac{\xi_{{\mathbf
k},\alpha}\xi_{\mathbf{ k+q},\beta}+\Delta_{{\mathbf
k},\alpha}\Delta_{\mathbf{ k+q},\beta}}{E_{{\mathbf
k},\alpha}E_{\mathbf{ k+q},\beta}})$, is in general nonzero for
$\alpha\neq\beta$ when ${\mathbf q}\rightarrow 0$. As a result,
$\chi^{a}_{L}(T)$ will in general be nonzero in the zero
temperature limit. However, the intraband contribution to the
orbital magnetic susceptibility should be suppressed in exact the
same manner as the spin susceptibility in the superconducting
state.

Now we present the numerical results for both the orbital and the
spin magnetic susceptibilities. In our calculation, the chemical
potential is self-consistently determined by solving the particle
number equation at each temperature. We then use Eq.(\ref{eqn5}) and
Eq.(\ref{eqn7}) to calculate the magnetic susceptibilities. The
temperature dependence of the pairing gap is modeled by
\begin{eqnarray}
\Delta(T)=\Delta\sqrt{1-\frac{T}{T_{c}}}, \nonumber
\end{eqnarray}
in which we have set $\Delta=0.02 eV$ and $k_{B}T_{c}=0.005 eV $.
These parameters are typical for the Iron-based superconductors. The
band filling will be fixed at $n=6$ at first, as the band structure
calculation leading to model Hamiltonian Eq.(\ref{eqn1}) is done at
such a commensurate filling.

\begin{figure}[h!]
\includegraphics[width=8cm,angle=0]{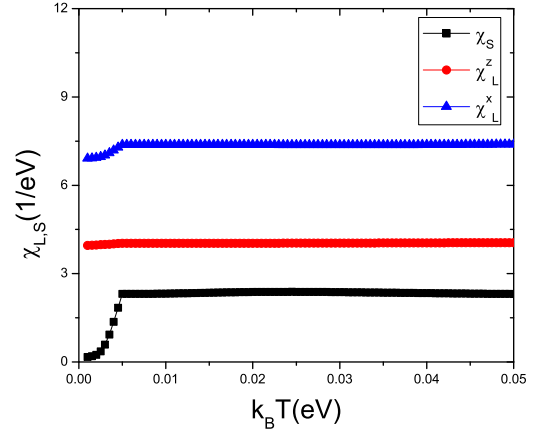}
\caption{The spin and orbital magnetic susceptibility of the
Iron-based superconductor calculated from the five-band model as
functions of temperature. The band filling is fixed at $n=6$.}
\label{fig2}
\end{figure}

The results for the orbital and spin magnetic susceptibilities are
shown in Fig.\ref{fig2}. Both susceptibilities are found to show
little temperature dependence below $k_{B}T=0.05eV$ in the normal
state. This is in accordance with the expectation for a typical band
metal. The orbital magnetic susceptibility is found to be several
time larger than the Pauli spin susceptibility and is already
comparable in magnitude with the observed total susceptibility. More
specifically, $\chi^{x}_{L}(T)$ is found to be about 3 times larger
than the Pauli spin susceptibility $\chi_{S}(T)$, which is estimated
to be about one-fourth of the observed total magnetic susceptibility
at 150 K\cite{Klingeler1111}. The orbital magnetic susceptibility
also show large anisotropy. The in-plane orbital susceptibility
$\chi^{x}_{L}(T)$ is found to be about 1.8 times larger than the
out-of-plane orbital susceptibility $\chi^{z}_{L}(T)$.

In our calculation, we have neglected the interaction correction.
The interaction correction is believed to induce enhancement of the
effective mass and thus the spin susceptibility. It is also believed
that the interaction correction will induce temperature dependence
in spin susceptibility. However, the interactions, such as the local
Coulomb repulsion and the Hund's rule coupling, are not expected to
renormalize the orbital magnetic susceptibility directly. Thus,
although the bare spin susceptibility calculated in this paper may
not be a reliable estimation for the real spin response, the results
for the orbital magnetic susceptibility should be robust.

In the superconducting state, the spin susceptibility $\chi_{S}(T)$
drops abruptly as the pairing gap develops on the Fermi surface and
approaches zero in the zero temperature limit. The in-plane orbital
magnetic susceptibility $\chi^{x}_{L}(T)$ also exhibits a
significant reduction in the superconducting state, but remains
nonzero in the zero temperature limit. On the other hand, the
signature of the superconducting transition in $\chi^{z}_{L}(T)$ is
almost unobservable. To understand why there is such a difference
between $\chi^{x}_{L}(T)$ and $\chi^{z}_{L}(T)$, we note that the
diagonal matrix element $O^{z}_{k,\alpha,\alpha}$ is identical zero
as the result of the tetragonal symmetry of the system, while
$O^{x}_{k,\alpha,\alpha}$ is in general nonzero. As we have
mentioned above, the intraband contribution to the orbital magnetic
susceptibility also suffers from a suppression by the coherence
factor in the superconducting state.

To see how our results depend on the band filling, we have carried
out the calculation for several different electron concentrations.
It is found that the qualitative features of both the spin and the
orbital magnetic susceptibilities, for example, their temperature
dependence and anisotropy, is quite robust, but their magnitudes are
reduced as we increase the electron concentration of the system. The
detailed dependence of the susceptibilities and their relative ratio
in the normal state on the band filling are shown in Fig.3. The
calculation is done at a fixed temperature $k_{B}T=0.03eV$ and the
rigid band approximation is assumed. It is found that both the
magnitude and the anisotropy of the orbital magnetic susceptibility
do not change significantly with band filling. On the other hand,
the spin susceptibility show much stronger band filling dependence.

\begin{figure}[h!]
\includegraphics[width=8cm,angle=0]{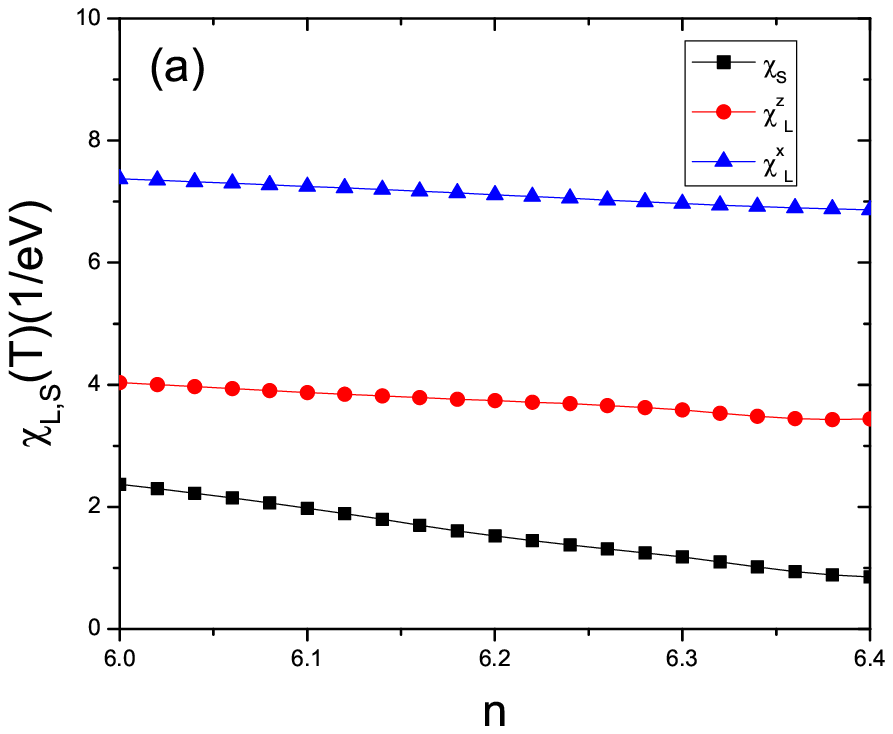}
\includegraphics[width=8cm,angle=0]{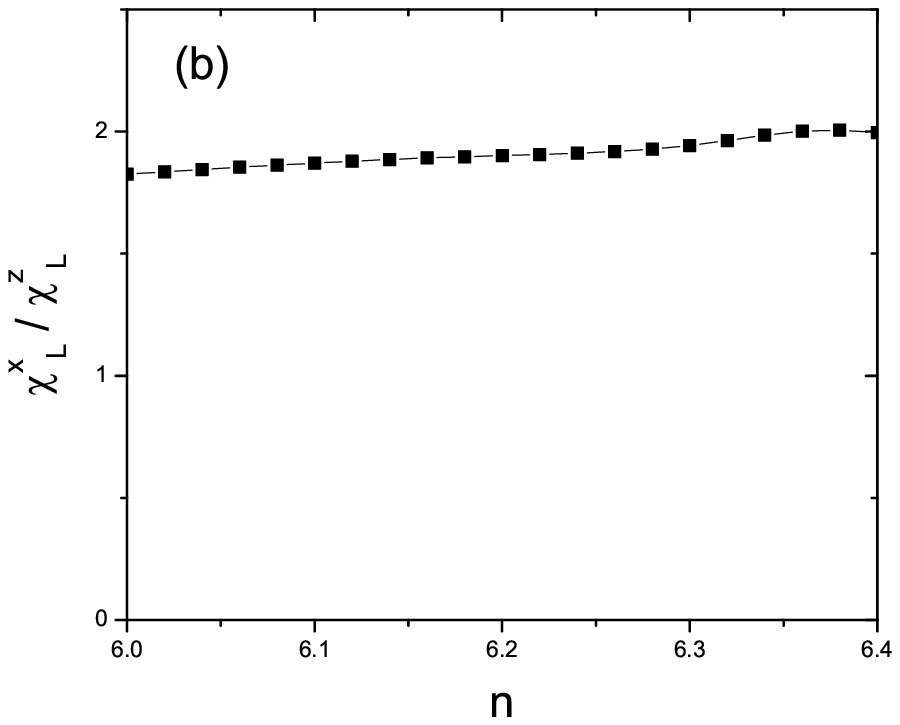}
\caption{(a) The spin and orbital magnetic susceptibility as
functions of the band filling at fixed temperature $k_{B}T=0.03eV$.
(b) The anisotropy ratio $\chi^{x}_{L}/\chi^{z}_{L}$ as a function
of band filling at fixed temperature $k_{B}T=0.03eV$.} \label{fig3}
\end{figure}

Now we turn back to the experimental data shown in Fig. \ref{fig1}.
As a common feature of all the data shown, both the in-plane and the
out-of-plane susceptibility exhibit linear temperature dependence
above $T_{c}(or\ T_{N})$ with almost the same slope. The in-plane
susceptibility is seen to be much larger than the out-of-plane
susceptibility in all measurements. Thus the total susceptibility
can be interpreted as consisting of two contributions: an isotropic
component with a linear temperature dependence and a temperature
independent component that is anisotropic. It is quite natural to
attribute the isotropic component to the spin magnetic response,
which with interaction correction can exhibit strong temperature
dependence\cite{GMZhang,SPKou,Chubukov}. The remaining anisotropic
component is more likely to be the contribution of the orbital
magnetic moment, which is not directly renormalized by the usual
local electron correlation effect and should be temperature
independent in the normal state. In principle, the anisotropy in
magnetic susceptibility can also be caused by the spin-orbital
coupling effect.\cite{Eremin} However, in the Iron-based
superconductors, the spin-orbital coupling is negligible small.
Therefore, we feel our proposal for the anisotropy is more realistic
for the Iron-based superconductors.

In summary, we have shown that the orbital angular momentum of the
conduction electrons in the Iron-based superconductors can play a
significant role in its low energy physics. It contributes a large
temperature-independent anisotropic component to the magnetic
susceptibility. As a result, the total magnetic response in the
normal state can be separated into a temperature dependent isotropic
part $\chi_{s}(T)$ and a temperature independent anisotropic part
$\chi_{L}(\theta)$, where $\theta$ is the angle between the magnetic
field and the normal of the Fe-As plane. In other words, a fit of
the experimental susceptibility to the formula
$\chi(T,\theta)=\chi_{s}(T)+\chi_{L}(\theta)$ should be feasible. We
note that the orbital magnetic moment can also contribute to the
relaxation of the nuclear spins and other fluctuation effect at low
energy. A full investigation of the dynamical orbital response will
be presented in future works.

YHS is support by NSFC Grant No. 10974167 and TL is supported by
NSFC Grant No. 10774187 and National Basic Research Program of
China No. 2007CB925001 and No. 2010CB923004.


\end{document}